\documentclass[doublecol]{epl2} 
\title{Statistical mechanics of two-dimensional foams}
\author{Marc Durand}
\institute{Mati\`{e}re et Syst\`{e}mes Complexes (MSC), UMR 7057 CNRS \& Universit\'{e}
Paris Diderot, 10 rue Alice Domon et L\'{e}onie Duquet, 75205 Paris Cedex 13, France, EU}
\usepackage{amsfonts}
\usepackage{amsmath}
\usepackage{amssymb}
\usepackage{graphicx}
\usepackage{rotating}
\usepackage{overpic}
\usepackage{tabularx}%
\usepackage{url}
\date{\today}

\abstract{
The methods of statistical mechanics are applied to two-dimensional foams under macroscopic agitation. A new variable - the total cell curvature - is introduced, which plays the role of energy in conventional statistical thermodynamics. 
The probability distribution of the number of sides for a cell of given area is derived. This expression allows to correlate the distribution of sides (``topological disorder") to the distribution of sizes (``geometrical disorder") in a foam. The model predictions agree well with available experimental data.}
\begin{document}
\maketitle

\section{Introduction}
Foams and related physical systems (like emulsions, biological tissues, or polycrystals) are ubiquitous, and serve as a paradigm for a wide range of physical phenomena and mathematical problems \cite{WeaireBook1,WeaireBook2,Morgan01,Morgan02}.
One of them deals with the general topological and geometrical properties of cellular materials \cite{Quilliet,Quilliet2,Delannay,Hilgenfeldt1,Hilgenfeldt2,Kafer,Hilgenfeldt3,Kraynik}.
In this connection, Quilliet \textit{et al.} \cite{Quilliet,Quilliet2} studied recently the topological features of two-dimensional (2D) soap froths under slow oscillatory shear. Such macroscopic strain induces rearrangements within the foam, and the number of sides of every cell evolves in time through local topological changes ($T1$ events) \cite{WeaireBook1,Cox1,Durand1}. Nevertheless, Quilliet \textit{et al.} reported the existence of an equilibrium state after few cycles, characterized by a stationary probability distribution of the number of sides per cell (topological disorder). They also showed that the width of this distribution of sides is strongly correlated to the distribution of bubble sizes (geometrical disorder) within the foam. 
These results suggest that the macroscopic state of a homogeneously sheared foam can be adequately described using the ideas and formalism of statistical thermodynamics.
Indeed, the pioneering work of Edwards on granular matter \cite{Edwards} has shown how the powerful arsenal of statistical physics can be extended to athermal systems. This method has proven its applicability to other fields as well \cite{divers}. 
Because there is no thermal averaging due to Brownian motion, this approach requires the presence of a macroscopic agitation (analogue to an effective temperature) allowing the system to explore its entire phase space.

Various attempts have been made in the past to describe the geometrical and topological properties
of 2D foams using the concepts of statistical thermodynamics \cite{Schliecker1,Schliecker2,Rivier1,Rivier2,Fortes3,Iglesias}.
However, these former theoretical approaches
rely on strong assumptions: either they use an ad-hoc interaction potential between bubbles \cite{Schliecker1,Schliecker2}, involve (rather than deduce) empirical laws correlating size and side distributions \cite{Rivier1,Rivier2,Fortes3}, or ignore some geometrical constraints (for instance, only the mean bubble area is specified, not the individual bubble areas \cite{Rivier1,Rivier2,Iglesias}). 
Some of these models are based on the maximum entropy (information theory) formalism \cite{Rivier1,Rivier2}, which has been subject to controversy \cite{Chiu}. Other models invoke minimisation of the energy \cite{Fortes3}, or a combination of both principles \cite{Iglesias} to describe the state of a foam. However, it has been established \cite{Weaire1, Graner1, Vaz1, Jiang, Kraynik} that different arrangements (topologies) of a large number of bubbles do not really affect the energy (see discussion below). 
Furthermore, none of these models can account for the correlations between topological and geometrical disorders reported by Quilliet \textit{et al.} \cite{Quilliet,Quilliet2}.

In this letter we set up a framework for describing the equilibrium state of a two-dimensional foam, basing our development on analogies with conventional statistical mechanics.
As for other athermal systems \cite{Edwards, divers}, we show that the energy is not relevant to describe the macroscopic state of a foam. Instead, a more appropriate state variable is introduced: the total cell curvature. 
We establish the function of state which is minimized for a finite cluster of bubbles at equilibrium. This thermodynamic potential function differs from entropy or energy used in previous theories.
The formalism developed here allows to derive an analytical expression for the distribution of the number of sides.
We show that the semi-empirical expression conjectured by
other authors \cite{Schliecker1,Schliecker2,Sherrington1,Sherrington2} is recovered, in a certain limit.
Finally, the bubble size-topology correlations deduced from the present theory are investigated, and compared with experimental data. 
\section{Physical, geometrical, and topological constraints}
Consider a given set of $N_{B}$ bubbles with prescribed areas $\lbrace A_{i} \rbrace$ (we focus on time scales much shorter than those typical of bubble coarsening and coalescence, so the bubbles preserve their integrity and size\footnote{Rigorously, the area and pressure of a cell vary from one configuration to another; only the number of molecules of gas it contains remains unchanged. We consider that the area fluctuations are small so that each bubble can be identified by its area.}). 
A 2D foam is a partition of the plane without gaps or overlaps, and its structure must obey certain constraints\cite{WeaireBook1} (one considers the dry foam limit where liquid volume fraction is negligible).
The physical constraints follow from the mechanical equilibrium throughout the system: 
first, the balance of film tensions at every vertex implies that the edges, or Plateau borders, are three-connected making angles of $120^{\circ}$ with each other (Plateau's laws).
Then, the balance of gas pressures in adjacent bubbles implies that every edge is an arc of circle, whose algebraic curvature $\kappa_{ij}$ is proportional to the pressure difference between the two adjacent bubbles $i$ and $j$ (Laplace's law):
\begin{equation}
\kappa_{ij}=-\kappa_{ji}=\dfrac{P_j-P_i}{\gamma},
\label{Laplace}
\end{equation}
where $\gamma$ is the film tension, 
and $P_i$ and $P_j$ are the pressures in bubble $i$ and $j$, respectively (by convention, $\kappa_{ij} \geq 0$ when the center of curvature is outside the cell $i$, i.e.: when $P_j \geq P_i$).
As a consequence, the algebraic curvatures of the three edges that meet at the same vertex must add to zero: 
\begin{equation}
\kappa_{ij}+\kappa_{jk}+\kappa_{ki}=0.
\label{Adjacent Laplace}
\end{equation}

The foam must also satisfy topological and geometrical requirements:
apart from the constraint of prescribed areas, its structure must obey Euler's rule, which relates the number of bubbles $N_B$, with those of Plateau borders $N_{Pb}$, and vertices $N_v$: $N_B+N_v-N_{Pb}=c$, where $c$ is the Euler-Poincar\'{e} characteristic ($c=0$ for a torus, $c=2$ for an infinite plane).
This rule, combined with Plateau's laws, immediately gives:
\begin{equation}
N_{Pb}=3(N_{B}-c).
\label{Euler-Plateau}
\end{equation}
Finally, the Gauss-Bonnet theorem applied to a $n$-sided cell yields $\sum_{j=1}^n l_{ij} \kappa_{ij}=\pi (n-6)/3$, where $l_{ij}$ denotes the length of the edge common to bubbles $i$ and $j$. 
\section{Microcanonical ensemble $\left(N_{B},\kappa_{tot},N_s \right)$}
In statistical mechanics, the most fundamental entry is certainly via the microcanonical ensemble. Suppose that the foam is agitated slowly as
compared to the characteristic relaxation time after a $T1$ process \cite{WeaireBook1,Cox1,Durand1} (one considers only perturbations that preserve the area of every bubble, and so the total foam area). Then, the deformation of the foam is quasistatic: its structure evolves through configurations which always satisfy the physical, geometrical and topological constraints stated above. Such configurations will be referred to as \textit{accessible microstates}. 

\begin{figure}[htb]
\begin{center}
\begin{overpic}[width=7.8cm]{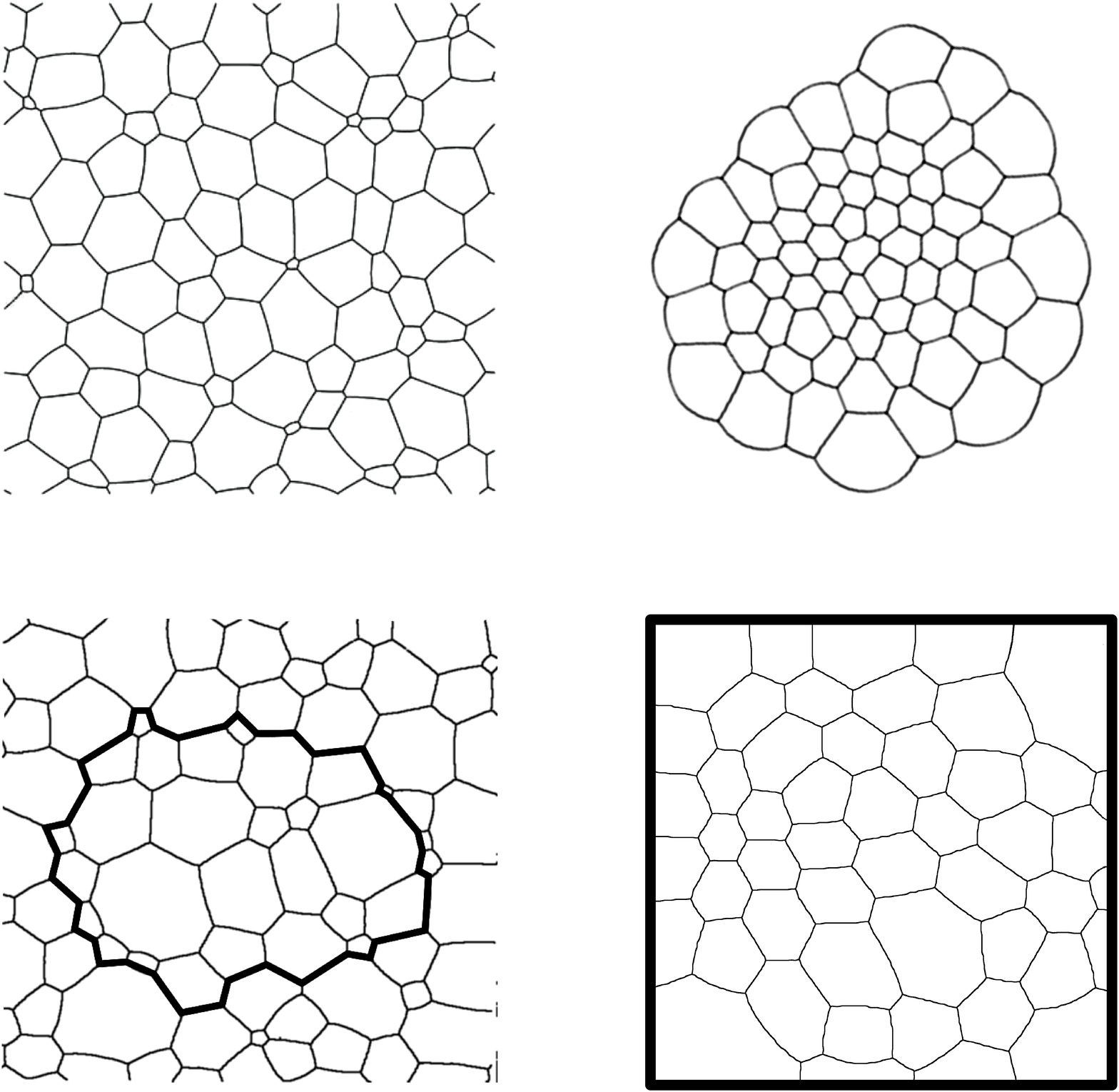}
\put(-6,55){\textbf{(a)}}
\put(50,55){\textbf{(b)}}
\put(-6,2){\textbf{(c)}}
\put(50,2){\textbf{(d)}}
\end{overpic}
\caption{Various situations for the boundary of a 2D foam. \textbf{(a):} infinite or periodic foam. \textbf{(b):} free cluster (figure taken from \cite{Graner1}). \textbf{(c):} cluster of bubbles within a larger foam. \textbf{(d):} foam enclosed in a container. Situation \textbf{(a)} is referred to as an unbounded foam, while situations \textbf{(c)} and \textbf{(d)} are referred to as bounded foams. Situation \textbf{(b)} is either an unbounded or bounded foam, depending on whether or not the surrounding air is included as a supplementary bubble.}
\label{boundary_conditions}%
\end{center}
\end{figure}

By analogy with the fundamental postulate of statistical mechanics \cite{Books}, we hypothesize that all the accessible microstates of a given set of bubbles filling the 2D space have equal probability, under a slow macroscopic agitation.
Obviously, a periodic or infinite foam fills the 2D space. It must be pointed out that the postulate applies to a free cluster too, provided that the surrounding air is included as a supplementary bubble. Then, this ``extra bubble" must also be taken into account in the counting of the number of accessible microstates. In the following, an infinite or periodic foam, or a free cluster plus the surrounding air shall be referred to as an \textit{unbounded foam}. Conversely, a free finite cluster, a cluster within a larger foam, or a foam enclosed in a container shall be referred to as a \textit{bounded foam}. Figure \ref{boundary_conditions} summarizes the different kinds of boundaries that exist for a 2D foam.
%

Rigorously, the accessible microstates do not all correspond to the same total surface energy. It is tempting, by analogy with the statistical mechanics of a gas, to restrict the equiprobability hypothesis to the microstates of equal energy. 
This refinement appears to be unnecessary: previous studies have shown that different arrangements of a large number of bubbles of given areas do not affect the energy very much\footnote{Actually, there is also some uncertainty on the value of the energy of an isolated volume of gas \cite{Books}.}: in their computational studies of 2D foams under shear, Jiang \textit{et al.} \cite{Jiang} reported energy fluctuations less than $2\%$. Graner \textit{et al.} \cite{Graner1} obtained the same results, both numerically and experimentally: the energy values of the different metastable states of a 2D foam lie within $2\%$ of the ``ground state" value. Kraynik \textit{et al.} \cite{Kraynik} conducted similar studies on 3D foams, and reported fluctuations below $4\%$ for the surface energy of every bubble.
These observations are also consistent with the so-called ``equation of state" of a foam \cite{Graner1,Ross,Aref,Fortes2}, which relates surface energy $E$, areas $A_i$, and pressures $P_i$ within a cluster of bubbles: 
$ E=\sum_{i}\left( P_{i}-P_{0}\right) A_{i} $, where $P_{0}$ is the external pressure. In the limit of fixed bubble pressures, the total surface energy of such a cluster would be conserved.
Hence, to a first approximation, one can reasonably assume that the energy of a large cluster does not depend on its configuration, but is directly determined from the number of bubbles $N_B$ and the area distribution $p(A)$.

%

\begin{figure}[htb]
\centering
\begin{overpic}[width=5cm]{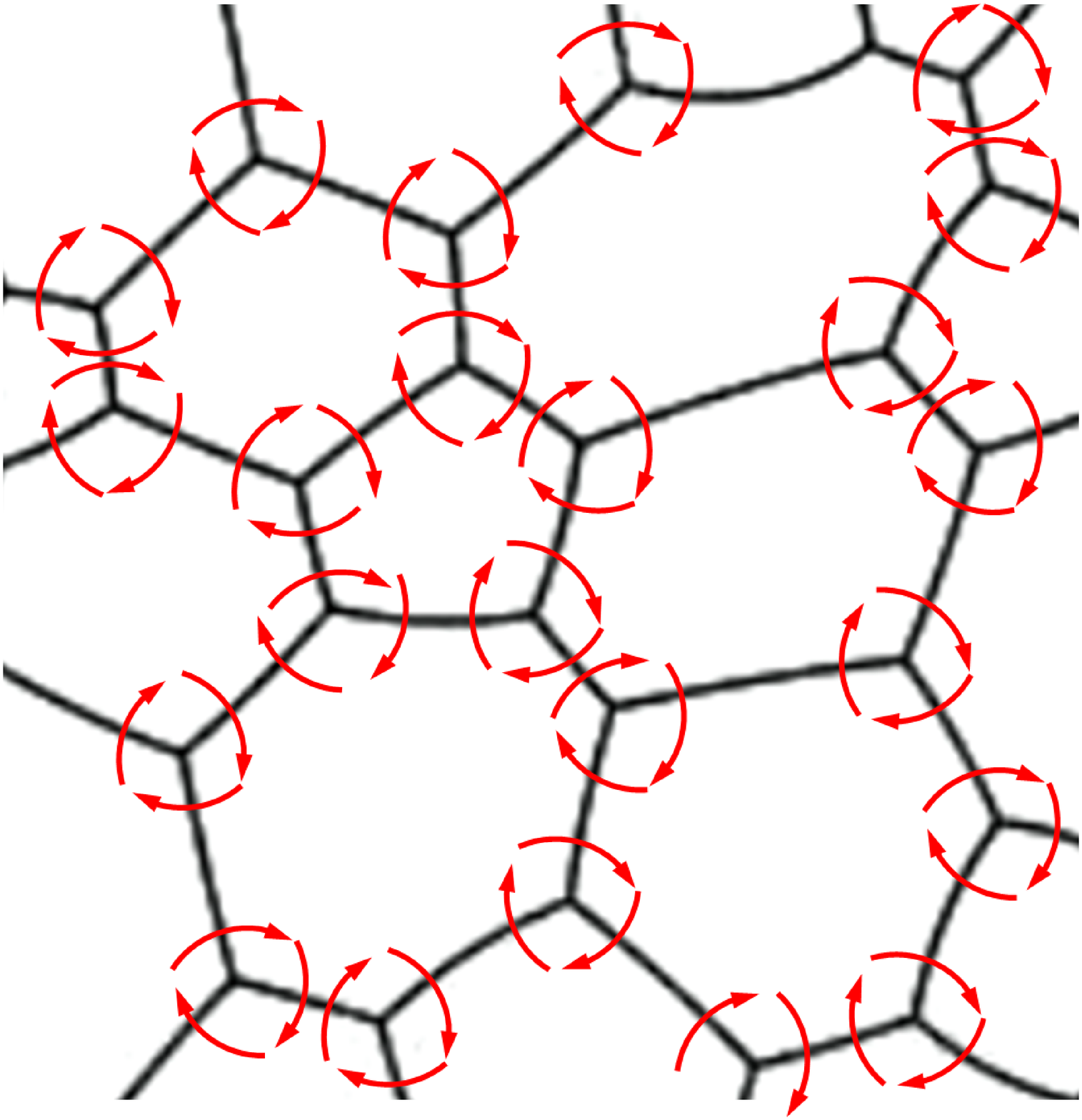}
\end{overpic}
\caption{Illustration of the equivalence (for an unbounded foam) between the sum (over all vertices) $\sum_\alpha S_\alpha$ and the sum (over all cells) $\kappa_{tot}=\sum_i \kappa_i$, where $S_\alpha$ denotes the sum -- turning clockwise -- of the algebraic curvatures $\kappa_{ij}+\kappa_{jk}+\kappa_{ki}$ between the three bubbles $i$, $j$, $k$ that share the same vertex $\alpha$, while $\kappa_i$ denotes the cell curvature of bubble $i$ (see Eq. (\ref{cell_curvature})). Each term $\kappa_{ij}$ in the first sum is represented by an arrow pointing from $i$ to $j$. The equivalence between the two sums is immediate: the cell curvature $\kappa_i$ of cell $i$ is represented by the collection of arrows coming out of that cell. For an unbounded foam at mechanical equilibrium, the curvature sum rule (\ref{Adjacent Laplace}) yields $S_\alpha=0$ at every vertex $\alpha$, and thus $\kappa_{tot}=0$. \label{sumrule}}
\end{figure}

On a macroscopic scale, the state of the foam should be described by a limited number of independent variables (besides the number of bubbles and the size distribution). These quantities must be ``constants of motion" \cite{Books}, i.e: they must keep constant values throughout the dynamics of the system (here, a slowly agitated foam). An important observation is that one can define two independent quantities which are conserved for an unbounded foam.
The first one is the \textit{total number of sides} $N_s=\sum_{i}n_{i}$, where $n_{i}$ is the number of sides of bubble $i$. One has $N_s=2N_{Pb}$ for an unbounded foam ($N_s$, and not $N_{Pb}$, is an extensive and fluctuating variable for a finite cluster). Thus, according to Eq. (\ref{Euler-Plateau}), $N_s=6(N_B-c)$.
The second one is the \textit{total cell curvature} $\kappa_{tot}= \sum_{i}\kappa_{i}$, where $\kappa_{i}$ is the \textit{cell curvature} of bubble $i$, defined as:
\begin{equation}
\kappa_{i}=\sum_{j \in \mathcal{N}(i)} \kappa_{ij},
\label{cell_curvature}
\end{equation} 
$\mathcal{N}(i)$ denoting the neighbouring cells of bubble $i$.
Obviously, $\kappa_{tot}$ is additive. 
Since $\kappa_{ij}=-\kappa_{ji}$, the terms cancel in pairs in the double sum, yielding $\kappa_{tot}=0$ for an unbounded foam.
The property $\kappa_{ij}=-\kappa_{ji}$ correlates the two adjacent bubbles that share a common edge. It is also important to note that -- regardless of this mathematical property -- the constraint $\kappa_{tot}=0$ is also imposed by the curvature sum rule (\ref{Adjacent Laplace}), as it is illustrated on Fig. \ref{sumrule}. This rule, which correlates the three adjacent bubbles that share a common vertex, has a physical origin (Laplace's law). 
Although it is unclear whether $N_s$ and $\kappa_{tot}$ are the only constants of motion for an unbounded foam, the constraints $N_s=6(N_B-c)$ and $\kappa_{tot}=0$ must be taken into account in the statistical description of such a system. Surprisingly enough, the constraint on the total curvature has always been ignored in previous theoretical models \cite{Schliecker1,Schliecker2,Rivier1,Rivier2,Fortes3,Iglesias}.

In the light of this discussion, we may restate the postulate as: 
all the accessible states of a given set of $N_B$ incompressible cells corresponding to the same values of $N_{s}$ and $\kappa_{tot}$ are equally probable, under macroscopic agitation. By analogy with thermal physics, the \textit{microcanonical ensemble} $\left(N_{B},\kappa_{tot},N_s \right)$ refers to a large number of copies of a foam (with given bubble size distribution p(A)) whose parameters $N_{B}$, $\kappa_{tot}$, and $N_s$ are fixed.
\section{Thermodynamic limit}
For a bounded foam, $\kappa_{tot}$ reduces to the sum of the side curvatures along its boundary. Both $N_s$ and $\kappa_{tot}$ are fluctuating variables, although their values are usually restricted within a certain range. For instance, for a free cluster, $N_s<2N_{Pb}$ and $\kappa_{tot}<0$ since the pressure in any bubble of the cluster is higher than the external pressure \cite{Vaz3}. 
For a cluster within a larger foam, $N_s<2N_{Pb}$, but $\kappa_{tot}$ can be positive or negative. For a foam enclosed in a container, $\kappa_{tot}=0$ (the container walls can be regarded as zero curvature sides), but $N_s$ fluctuates.
One can argue whether or not the fluctuations of $N_s$ and $\kappa_{tot}$ become negligible in the thermodynamic limit ($N_B \rightarrow \infty$, total surface area $A_{tot} \rightarrow \infty$, but $\langle A \rangle=A_{tot}/N_B$ remains finite).
The average number of sides of a large cluster ($N_B \gg 1$) scales as $\langle N_{s}\rangle \sim N_B$, while the total curvature, equal to the sum of the side curvatures along the cluster boundary, scales as $\langle \kappa_{tot} \rangle \sim \sqrt{N_B/\langle A \rangle}$.
The relative standard deviations of such quantities scale as $N_B^{-1/2}$ \cite{Books}. Thus, the fluctuations of $N_s$ and $\kappa_{tot}$ become vanishingly small as $N_B \rightarrow \infty$. Moreover, their thermodynamic limit values are known: $\left\langle N_s \right\rangle / N_B \rightarrow 6$, and $\left\langle \kappa_{tot} \right\rangle / N_B \rightarrow 0$.
Note however that $\langle N^{*}_{s}\rangle/ N^{*}_B \sim {N^{*}_B}^0$ converges much faster than $\langle \kappa_{tot}^{*} \rangle /N^{*}_B \sim {N^{*}_B}^{-1/2}$.
\section{Idealized foam}
In order to obtain the probability distribution of the number of sides for a cell of given size, we need to enumerate the microstates which correspond to the same macroscopic state. Ideally, a microstate is specified by the curvature, length, position and orientation of each Plateau border.
However, this description is too cumbersome to handle. Moreover, even for a given foam topology, the number of different accessible microstates depend in a non-trivial way on boundary conditions (e.g.: free, periodic, or enclosed cluster) and bubble size distribution \cite{Herdtle, Fortes1, Vaz2, Graner1, Weaire2}. 
\begin{figure}[htb]
\begin{center}
\begin{overpic}[width=7.8cm]{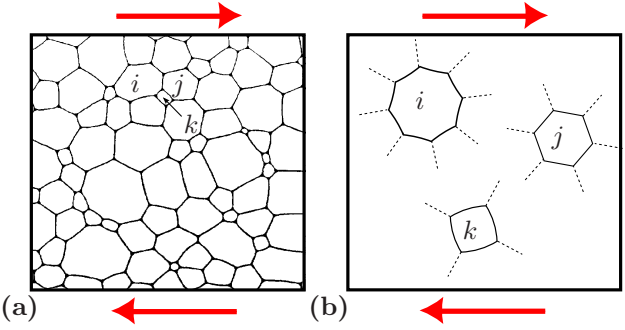}
\put(17,39){$i$}
\put(24,39){$j$}
\put(26,32){$k$}
\put(65,36){$i$}
\put(88,30){$j$}
\put(73,14){$k$}
\put(-5,1){\textbf{(a)}}
\put(47,1){\textbf{(b)}}
\end{overpic}
\caption{\textbf{(a):} Real two-dimensional foam under shear. \textbf{(b):} Idealization of the foam (mean-field approximation): adjacent bubbles $i$, $j$ and $k$ are now regular cells surrounded by an effective foam and disconnected from each other.\label{2dfoams}}
\end{center}
\end{figure}

We shall use a simplified microscopic description of the foam: 
we consider that for each particular bubble $i$, the rest of the foam can be replaced
on average by a ``mean field" of identical bubbles. Thus, the bubble $i$ is a regular cell with identical curved sides joining at each vertex with angles of $120^\circ$, as sketched on Fig. \ref{2dfoams}b.
The cell curvature of a $n$-sided regular cell with area $A$ is \cite{Iglesias,Graner1}:
\begin{equation}
\kappa\left(A,n \right) =\dfrac{\pi}{3}\dfrac{n \left(n-6 \right)}{e(n)\sqrt{A}},
\end{equation}
where $e(n)=\frac{\pi \mid n-6 \mid}{3 \sqrt{n}} \left( \frac{\pi}{n} - \frac{\pi}{6} - \frac{\sin (\pi/n - \pi/6) \sin(\pi/6)}{\sin(\pi/n)} \right)^{-1/2}$ is the elongation of the cell (ratio of perimeter to square-root of area). $e(n)$ is a slowly decreasing function, lying between $e(3)\simeq 3.74$ and $e(\infty)\simeq 3.71$ \cite{Graner1}. 
It can be noticed that while the (surface) energy of such a bubble ($\sim e(n)\sqrt{A}$) is almost independent of $n$, its cell curvature $\kappa\left(A,n \right)$ increases rapidly with $n$, and is not upper-bounded.

In this idealized foam description, each bubble is ``disconnected" from the others.
By construction, all the physical, geometrical and topological constraints are satisfied, except the curvature sum rule (\ref{Adjacent Laplace}) and the Euler-Plateau relation (\ref{Euler-Plateau}), which both involve adjacency of the bubbles.
Consistent with the discussion above, the \textit{accessible} microstates of a foam tiling the entire plane are those which satisfy $N_s=6(N_B-c)$ and $\kappa_{tot}=0$ (the number of bubbles is large enough so $N_s$ and $\kappa_{tot}$ can be treated as continuous variables).
A microstate of the idealized foam is specified by the numbers of sides $\left\lbrace n_i \right\rbrace $ of all the bubbles.
Hence, the number of accessible microstates is obtained by enumerating the distributions of the $N_s=6(N_B-c)$ sides over the $N_{B}$ bubbles which satisfy the constraint $\kappa_{tot}=0$. This number is not easy to evaluate, and a grand-canonical description shall be more appropriate.
\section{Grand-canonical ensemble $\left(N_{B},\beta^{-1},\mu \right)$}
Consider then a sample of $N^{*}_B$ bubbles in the unbounded foam (the asterisk denotes the variables of this grand-canonical ensemble). This sample can be a cluster of bubbles, or a collection of isolated bubbles. Let $p^{*}(A)$ be the distribution of bubble size within this sample.
Although $N^{*}_B$ is fixed, this system can exchange sides and curvature with the rest of the foam, through $T1$ events. Hence, the total number of sides $N^{*}_s$ and the total curvature $\kappa^{*}_{tot}$ are now internal variables free to fluctuate for this system\footnote{$N^{*}_s$ and $\kappa^{*}_{tot}$ are independent variables: there are different ways of distributing $N^{*}_s$ sides over $N^{*}_B$ bubbles; each of these distributions corresponds to a different value of $\kappa^{*}_{tot}$.}.
We assume that possible other variables required to describe the macroscopic state remain fixed. Surface energy in particular does not fluctuate, since it depends only on $N^{*}_B$ and the distribution $p^{*}(A)$, and not on the specific configuration of the system.
Suppose that the rest of the foam is large in comparison with the system, so that it constitutes a reservoir of sides and curvature.
Using the formalism of conventional statistical mechanics \cite{Books}, it comes that the probability for the system to be in the microstate $\left( n_1,n_2,\ldots,n_{N^{*}_{B}} \right)$
is proportional to $e^{-\beta \kappa^{*}_{tot} +\mu N^{*}_s}$, 
with $\kappa^{*}_{tot}=\sum_{i=1}^{N^{*}_B} \kappa(A_i,n_i)$
and $N^{*}_s=\sum_{i=1}^{N^{*}_B} n_{i}$. 
$\beta^{-1}$ and $\mu$ (rigorously, $\mu \beta^{-1}$) denote respectively the ``temperature" of the reservoir of curvature, and the ``chemical potential" of the reservoir of sides. A large number of copies of a foam whose parameters $\left(N_{B},\beta^{-1},\mu \right)$ are fixed shall be referred to as a \textit{grand-canonical ensemble}.
As noticed before, the cell curvature of a bubble is not upper-bounded as $n \rightarrow \infty$, what ensures that the temperature $\beta^{-1}$ is always positive \cite{Books}.
Finally, the probability for a given cell of size $A$ to have $n$ sides is: 
\begin{equation}
p_{A}(n)=\chi^{-1}(A) e^{-\beta \kappa(A,n) +\mu n},
\label{distribution1}
\end{equation}
where $\chi(A)= \sum_{n \geq 3} e^{-\beta \kappa(A,n) +\mu n}$ denotes the partition function of the cell.
The average total cell curvature and average number of sides are, respectively,
$\langle \kappa^{*}_{tot}\rangle=-\partial \ln \Xi / \partial \beta$
and $\left\langle N^{*}_{s}\right\rangle=\partial \ln \Xi / \partial \mu$, 
with $\ln \Xi=N^{*}_B \int_{0}^{\infty}p^{*}(A) \ln \chi(A) dA$. $\Xi$ is the partition function of the system.
Using the formalism of conventional statistical mechanics \cite{Books}, one concludes that the thermodynamic equilibrium of a foam in contact with a reservoir of curvature and sides coincides with the minimum of the thermodynamic potential $\Phi=-\beta^{-1}\ln \Xi$.
The analogy between the statistical descriptions of an ideal gas and a 2D foam is summarized in Table \ref{tableau}.

As noted by Graner and coworkers \cite{Graner1}, the elongation of a regular cell is almost constant: $e(n)\simeq \overline{e} \simeq 3.72$. With this simplification, the total surface energy of the cluster is strictly conserved (i.e., it does not depend on the cluster configuration). Besides, the distribution (\ref{distribution1}) simplifies to 
\begin{equation}
p_{A}(n)=\chi'^{-1}\left( A \right)  e^{-\beta'(n-6)^2 +\mu'(n-6)},
\label{distribution2}
\end{equation}
with $\chi'=\chi e^{-6 \mu}$,
$\beta'(A)=\pi \beta/(3 \overline{e} \sqrt{A})$
and
$\mu'(A)=\mu-2 \pi \beta/( \overline{e} \sqrt{A})$. This is the exact distribution intuited independently by 
Schliecker and Klapp \cite{Schliecker1,Schliecker2} and Sherrington and coworkers \cite{Sherrington1,Sherrington2}, except that here $\beta'$ and $\mu'$ explicitly depend on the bubble size $A$. Such a dependence 
is necessary to reflect 
the correlations between bubble size and bubble shape which have been observed experimentally \cite{Quilliet,Quilliet2}.
It can be noted that the average number of sides $\overline{n}(A)=\sum_{n\geqslant3}np_A(n)$ increases with $A$. This result is consistent with experimental observations \cite{WeaireBook1, Glazier}: larger bubbles have more sides since they are surrounded by smaller bubbles.
%
\begin{table}
\caption{Statistical description of an ideal gas and a 2D foam.}%
\label{tableau}%
\centering
\newcolumntype{Y}{>{\centering\arraybackslash}X}
\begin{tabularx}{\linewidth}[c]{|Y|Y|Y|}\hline
& \textbf{ideal gas} & \textbf{2D foam}\\ \hline
\textbf{source of ergodicity} & Brownian motion / collisions & macroscopic shear / T1 events \\ \hline
\textbf{constants} & masses $\lbrace m_i\rbrace$  & areas $\lbrace A_i\rbrace$ \\\hline
\textbf{degrees of freedom} & positions and momenta $\lbrace\mathbf{r}_i, \mathbf{p}_i\rbrace$& side numbers $\lbrace n_i\rbrace$ \\\hline
\textbf{fixed parameters (microcanonical ensemble)} & energy $E$, number of molecules $N$, volume $V$  & total curvature $\kappa_{tot}$, number of sides $N_s$, number of bubbles $N_B$ \\\hline
\textbf{fixed parameters (grand-canonical ensemble)} & temperature $T$, chemical potential $\mu$, volume $V$ & effective temperature $\beta^{-1}$, effective chemical potential $\mu$, number of bubbles $N_B$ \\\hline
\end{tabularx}
\end{table}
%
%
%
%

It must be pointed out that the present theory contains no free parameters in the thermodynamic limit: $\left\langle N_s^{*} \right\rangle / N^{*}_B$ and $\left\langle \kappa_{tot}^{*} \right\rangle / N^{*}_B$ have known values, and $\beta$ and $\mu$ are directly obtained by extremizing the grand-canonical entropy per bubble 
\begin{equation}
S(\beta,\mu)=\int_{0}^{\infty}p^{*}(A) \ln \chi(A) \mathrm{d}A+ \beta \dfrac{\left\langle \kappa_{tot}^{*} \right\rangle}{N^{*}_B}-\mu \dfrac{\left\langle N_s^{*} \right\rangle}{N^{*}_B},
\end{equation}
with $\left\langle N_s^{*} \right\rangle / N^{*}_B = 6$ and $\left\langle \kappa_{tot}^{*} \right\rangle / N^{*}_B = 0$. 
\section{Discussion} 
The expression (\ref{distribution2}) allows to deduce the distribution of number of sides per cell $p(n)$ from the distribution of bubble size $p(A)$ [$p(n)=\int_0^\infty p(A)p_A(n)\mathrm{d}A$], and thus to correlate topological and geometrical disorders. Let us study the implications of the present theory with the simple case of an infinite/periodic monodisperse foam: 
$p(A)=\delta(A-A_0)$. In that case, the thermodynamic limit values $\langle N^{*}_{s}\rangle/N^{*}_{B}=6$ and $\langle\kappa^{*}_{tot}\rangle/N^{*}_{B}=0$ give, respectively, $\langle n-6 \rangle=0$ and $\langle(n-6)^2\rangle=0$. Thus, the distribution of side number tends to the Kronecker delta distribution: $p(n)=p_{A_{0}}(n)=\delta_{n,6}$. As expected, all the bubbles of an unbounded monodisperse foam have hexagonal shape.

\begin{figure}[htb]
\centering
\begin{overpic}[width=8cm]{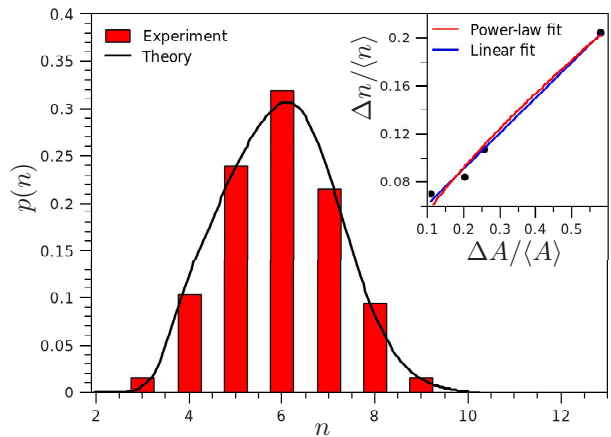}
\put(50,3){$n$}
\put(0,40){\begin{sideways} $p(n)$ \end{sideways}}
\put(75,32){$ \Delta A / \langle A \rangle $}
\put(56,55){\begin{sideways} $\Delta n / \langle n \rangle$ \end{sideways}}
\end{overpic}
\caption{Theoretical and experimental distributions $p(n)$. The experimental distribution is taken from \cite{Quilliet}. The theoretical distribution is obtained by fitting the function $p(n)=\int_0^\infty p(A)p_A(n)\mathrm{d}A$ to the data. Inset: topological disorder $\Delta n / \langle n \rangle$ vs geometrical disorder $ \Delta A / \langle A \rangle $ for four different foams. Black dots: obtained from theory; red curve: power-law fit $a \left( \Delta A / \langle A \rangle \right) ^b$, with $a=0.30\pm 0.03$ and $b=0.74\pm 0.08$; blue curve: linear fit $a\Delta A / \langle A \rangle+b$, with $a=0.30\pm 0.02$ and $b=0.03\pm 0.01$.}
\label{distribution-b}
\end{figure}
We also compare the model with the four experimental distributions reported in the Quilliet \textit{et al.} paper \cite{Quilliet}. 
Foam samples contain a few hundreds of bubbles. 
This number is large enough to consider that $\langle N^{*}_{s}\rangle /N^{*}_B \simeq 6$, but not large enough to assume that $\langle \kappa_{tot}^{*} \rangle \sqrt{\langle A^{*} \rangle} /N^{*}_B \simeq 0$. Since the finite value of $\langle \kappa_{tot}^{*} \rangle /N^{*}_B$ is unknown,
$\beta$ and $\mu$ cannot be obtained by extremizing $S(\beta,\mu)$. Instead, they are obtained by fitting the theoretical distribution of sides $p(n)=\int_0^\infty p(A)p_A(n)\mathrm{d}A$ -- in which $p(A)$ is the experimental size distribution, and $p_A(n)$ is given by Eq. (\ref{distribution2}) -- to the data.  Regression is performed under the constraint $\sum_{n \geq 3}np(n)=6$, so there is only one adjustable parameter really. 
Fig. \ref{distribution-b} compares a theoretical distribution $p(n)$ obtained this way with the corresponding experimental distribution taken from \cite{Quilliet}.
The theoretical curve reproduces the experimental data well. Comparison of the model with the other distributions of \cite{Quilliet} (not shown here) gives similar results.
We also plotted the relative standard deviation $\Delta n/\langle n \rangle=\sqrt{\langle n^2 \rangle - \langle n \rangle^2}/\langle n \rangle$ of the theoretical distributions, as a function of the relative standard deviation $\Delta A/\langle A \rangle$ of the size distributions for the four samples (inset of Fig. \ref{distribution-b}). 
A power-law fit of this plot is in good agreement with the relationship found by Quillet \textit{et al.}: $\Delta n/\langle n \rangle=0.27 \left( \Delta A/\langle A \rangle\right)^{0.8}$. However, it can be noticed that the power-law fit can be hardly distinguished from a linear fit. 
%

In summary, we developed a theoretical model to describe the state of a two-dimensional foam under slow agitation, using a formulation closer to conventional statistical mechanics than information theory. We show that the total number of sides and the total cell curvature -- rather than energy -- are the relevant variables to describe the macroscopic state of a foam. The distribution of sides of a cell of given size is derived. This result allows to correlate the size and shape distributions. Theoretical size-topology relations deduced from the theory are in very good agreement with existing experimental data. However, a free parameter has been added for the comparison with experiments, due to the indeterminacy of the experimental values of $\langle \kappa_{tot}^{*} \rangle /N^{*}_B$. Furthermore, the grand-canonical description requires that $1\ll N^{*}_B \ll N_B$. This condition cannot be checked on the available data. Further experiments taking these considerations into account should allow to confirm the validity of the present (zero-free-parameter) model.

The formalism developed here can be extended to three-dimensional foams, and to coarsening (and coalescing) foams. Coarsening has consequences for the size-topology correlations particularly, since the average area of $n$-sided cells increases more rapidly in that case \cite{Quilliet2,Glazier,Flyvbjerg}. Extension of this formalism to other cellular systems (concentrated emulsions, cell aggregates) is also under investigation.

\acknowledgments
I wish to thank J. B. Fournier, F. Graner and F. Van Wijland for useful discussions and suggestions.

\end{document}